\documentclass[aps,showpacs,letterpaper,twocolumn,prb,reprint,floatfix,natbib,longbibliography]{revtex4-1}
\usepackage{graphicx}
\usepackage{dcolumn}
\usepackage{bm}
\usepackage{hyperref}
\usepackage{amssymb}
\usepackage{amsmath,mathtools}
\usepackage{esint}
\usepackage[caption=false]{subfig}
\usepackage{epstopdf}
\usepackage{amsfonts}
\usepackage{color}

\DeclareMathOperator\ext{ext}
\newcounter{subeqn} %
\makeatletter

\begin{document}
\title{Geometric effects 
in non-equilibrium electron transfer statistics in adiabatically driven quantum junctions}
\author{Himangshu Prabal Goswami$^{1}$}
\author{Bijay Kumar Agarwalla$^{2}$}
\author{Upendra Harbola$^{1}$}
\affiliation{$^1$Department of Inorganic and Physical Chemistry, Indian Institute
of Science, Bangalore-560012, India}
\affiliation{$^{2}$Chemical Physics Theory Group, Department of Chemistry, and Centre for Quantum Information and 
Quantum Control, University of Toronto, 80 Saint George St., Toronto, Ontario, Canada M5S 3H6}
\date{\today}
\begin{abstract}
Cyclic Pancharatnam-Berry (PB) and adiabatic noncyclic 
geometric (ANG) effects are investigated in  a single electron orbital system connected to two 
metal
 contacts with externally driven
  chemical potential and/or temperatures. 
  The PB contribution doesn't affect the
 density matrix evolution, but has quantitative effect on the 
 statistics (fluctuations) of electron
  transfer. The ANG contribution, on the other hand, affects the net flux across the junction.
  Unlike the PB, the ANG contribution is non-zero when 
   two parameters are identically driven.
  Closed analytical expressions are derived for the ANG contribution to the flux,
  and the PB contribution to the
   first two leading order fluctuations. Fluctuations can be modified by manipulating
    the relative phases of the drivings. Interestingly, 
    we find that the fluctuations of the pumped charge do not satisfy the steady state fluctuation theorem
    in presence of  nonzero geometric contribution, but can be recovered for a vanishing geometric contribution
     even in presence of the external driving.
    
\end{abstract}
\pacs{05.60.Gg,05.70.Ln,72.10.Bg,03.65.Vf}

\maketitle
\section{Introduction}
A parametric modulation of a system Hamiltonian in an adiabatic fashion adds a
phase change in system
state. This phase results from holonomy of the parameter space
and is known as the geometric phase \cite{griffiths2005introduction,bohm2013geometric}. 
The geometric phase is 
quantified by the area traced in the parameter space. At least two independent parameters in the Hamiltonian
should be subjected to time modulation.  When the  parametrization is cyclic, it
is commonly referred to as the Pancharatnam-Berry (PB) phase\cite{pancharatnam1956generalized,berry1984quantal}.
For noncyclic evolution of the parameters, 
 the acquired phase is known as adiabatic
 noncyclic geometric (ANG) phase\cite{pati,PhysRevLett.60.2339}. The geometric phases
 realized in systems with no degeneracy in the eigenspace are usually referred to as abelian.
 In presence of degeneracy, the holonomies
  do not commute
and  give rise to non-abelian geometric phases
\cite{anandan1988non,aharonov1987phase,budich2012all,duan2001geometric}.

Over the years, the effect of geometric phase has been studied in several 
systems such as solids\cite{taguchi2001spin}, condensed matter systems\cite{zhang2005experimental} and
quantum qubits \cite{leek2007observation} which can
affect both the
physical \cite{xiao2010berry} and chemical\cite{lu2010blowing} properties
of quantum systems.
There have been several attempts to understand
 the role of geometric phases in open quantum systems
 \cite{carollo2003geometric,sarandy2006abelian,calvani2014open,ren2010berry,hu2014berry,PhysRevLett.Ross}.
 However due to the complexity
  of non-equilibrium quantum systems, the exact role of geometric phase remains a mystery. 
  A special type of non-equilibrium systems are quantum junctions where
  heat and electron transport
   are the key dynamic processes. These systems are made up
  of a quantum system which is 
  coupled to reservoirs (bosonic or fermionic) at different thermodynamic states. Such junctions
    are realized in molecular break junctions\cite{reed1997conductance,huang2015break}, quantum heat engines
    \cite{kosloff2013quantum,goswami2013thermodynamics}, 
    single molecules sandwiched
     between a Scanning Tunneling Microscope (STM) 
     tip and metal surface\cite{reecht2014electroluminescence,latha}
     as well as single molecule electronic devices  \cite{song2011single,sun2014single,fuzi}. 
     Most of the literature on quantum junctions focuses on the steady-state transport properties.
     In recent years it has been recognized that a time
      dependent probe (optical or electronic) coupled to transport measurements can lead to better
      characterization of the junction 
      \cite{harbola2014frequency,goswami2015electroluminescence,shamai,vahid}. Such probes are usually
       fast (comparable to relaxation of the system). Here we explore the other extreme when the external
        driving is slow and the geometric phase is well defined.
    The geometric phase
    is known to create charge pumping in open quantum dots \cite{brower,switkes} and 
     interference in spin currents through single-molecule 
    magnets \cite{smm}. 
     However its role in affecting electron transfer statistics
    has not been  explored so far, although theoretical methods exist to explore the dynamics and
      statistics
   of such non-equilibrium quantum systems \cite{uhpr,dharheat,ward,nazarov,nit,uhrmp}.

    In this work, we study the geometric effects due to
    cyclic (PB) and noncyclic (ANG) adiabatic parametrization
    on electron transfer statistics 
    induced by time-dependent adiabatic change in 
 thermodynamic equilibrium of electronic leads. We formulate a general theory 
 of the geometric effects for non-equilibrium electron transport in weakly coupled systems
 and apply it to a single resonant level model. 
 We find that the time-evolution of the reduced density matrix 
  is not affected by the PB contribution but leads to a quantitative change in the statistics. 
   On the other hand, the ANG part globally affects the dynamics.  
  The PB contribution
  can be manipulated to alter the electron transfer statistics from antibunched to bunched
  by manipulating the phase-difference between the two drivings. 
   It was recently shown that in case of bosonic reservoirs coupled to quantum system,
pumping was possible by modulating the reservoir temperatures periodically\cite{ren2010berry}. 
We however find that for electron transfer, PB contribution does not affect  
   the average electronic flux. The ANG part however affects the evolution of the density
    matrix and therefore contributes to the total electronic flux.
     In presence of geometric contributions (cyclic or noncyclic), fluctuations
 in the electrons exchanged between leads 
 do not satisfy the standard fluctuation theorem (FT),
  $\lim_{t\rightarrow\infty}\ln [P(q,t)/P(-q,t)]=q\mathfrak F$\cite{uhrmp}, where
  $P(q,t)$ is the probability distribution function (PDF) for the
  net number ($q$) of electrons exchanged 
  between leads in a measurement time $t$,
  and  $\mathfrak F$ is the thermodynamic force associated to the electron-flux.
  Similar violation of the 
  fluctuation theorem or the Gallovoti-Cohen (GC) symmetry \cite{gal,dhar2} was also
   reported in case of heat transport \cite{ren2010berry}. However, unlike  
   results of Ren et al\cite{ren2010berry}, for a vanishing 
    geometric contribution, we recover the steady state FT for any 
    phase difference between the drivings.

The paper is organized as follows. In Sec.\ref{gc}, we formulate a general interpretation of
geometric contributions in 
weakly coupled quantum junction using Liouville space formalism \cite{uhpr}. 
In Sec. \ref{srl}, we present a driven quantum master equation (QME) for a single level system and analytically
evaluate the PB contribution that arises due to a periodic time
modulation of the thermodynamic equilibrium of the electronic reservoirs. In 
Sec. \ref{stats}, we discuss the effect of PB contribution in electron transfer statistics. In Sec.\ref{SSFT}, we show the
violation of the steady-state fluctuation theorem, which is recovered
 for a vanishing geometric part. In Sec. \ref{NCE} we discuss
 the effect of ANG contribution on the dynamics of electron transport and then we 
 conclude in Sec.\ref{conc}.

\section{ The geometric curvature}
\label{gc}
  Geometric effects can be realized in systems with adiabatic 
  external driving. A weakly coupled system dynamics in the 
  Liouville space is governed by the quantum-Liouville equation
   \cite{njp,PhysRevA.93.032118},
\begin{equation}
\label{qLiouville}
 |\dot\rho(t)\rangle\rangle=\hat{\mathcal L}(t)|\rho(t)\rangle\rangle,
\end{equation}
where $|\dot\rho(t)\rangle\rangle$ is the time rate of change of the reduced density vector for the
system.
$\hat{\mathcal L}(t)$ is the Liouvillian superoperator containing the time dependent driving. 
Equation (\ref{qLiouville}) is valid for systems with no degenerate energy levels
that get mixed due to the interaction with the baths (leads). The lead correlations 
decay much faster
 than the system relaxation.
We assume that $\hat{\mathcal L}(t)$ is diagonalizable, has 
 a single zero eigenvalue to guarantee the existence of a well defined steady state, and contains
  well-separated eigenvalues. 
\begin{equation}
\Lambda(t)=U^{-1}(t) \hat{\mathcal L}(t)U(t).
\end{equation}
Here, $U(t)(U^{-1}(t))$ is a matrix composed of the instantaneous 
right (left) eigenvectors of $\hat{\mathcal L}(t)$ and diagonalizes
 $\hat{\mathcal L}(t)$ to $\Lambda(t)$. This defines a new basis (eigenbasis of $\hat{\mathcal L}$), where
 \begin{equation}
 \label{eigenbasis}
 |\varrho(t)\rangle\rangle=U(t)^{-1} |\rho(t)\rangle\rangle.
 \end{equation}
Equation of motion in this new basis is
\begin{eqnarray}
|\dot\varrho(t)\rangle\rangle&=&[\Lambda(t)-U^{-1}(t)\dot U(t)]|\varrho(t)\rangle\rangle.
\label{der-U}
\end{eqnarray}
In the adiabatic limit, the external driving is  assumed to be
much slower as compared to the internal system relaxations
such that there are no transitions between the eigen states.
This amounts to neglecting
the off-diagonal terms of $U^{-1}(t)\dot U(t)$ in Eq.(\ref{der-U}).
We denote $B_d(t)=\text{diag}\{U^{-1}(t)\dot U(t)\}$ and write the 
solution of Eq. (\ref{der-U}) as
\begin{eqnarray}
\label{5}
 |\varrho(t)\rangle\rangle&=&e^{\int_0^t dt'\Lambda(t')-\int_0^t dt'B_d(t')}|\varrho(0)\rangle\rangle.
\end{eqnarray}
 The adiabatic approximation is considered only for the system evolution.
The system evolution is adiabatic with respect its relaxation to the steady
state determined by the coupling to the leads. There are three time scales: bath relaxation ($\tau_B$),
 system relaxation ($\tau_s$) due to coupling with the bath, and the time 
  period ($t_p$) of the external driving. The present formulation assumes 
  that $\tau_B\ll\tau_s\ll t_p$. 
Faster the relaxation of the lead correlations, better is the approximation.

Using Eqs. (\ref{eigenbasis}) and (\ref{5}), we get,
\begin{equation}
\label{berryEq}
 |\rho(t)\rangle\rangle=U(t)e^{\int_0^t dt'\Lambda(t')}e^{-\int_0^tdt'B_d(t')}U^{-1}(0)|\rho(0)\rangle\rangle.
\end{equation}
The first exponential is the usual 'dynamic' contribution to the time evolution. The second
 exponential is an additional part acquired due to the external driving and has a geometric
  interpretation.
In the absence of driving, $B_d=0$. 
 Let $\textbf{x}$ 
 represent a vector space corresponding to any two parameters, say $x$ and $y$, 
 that are being modulated externally and
  periodically
  in time.
 We can then convert the time integral of the 
 second exponential in  Eq.(\ref{berryEq}) to a line integral along a contour
  $\mathcal C$, representing the instantaneous $x$ and $y$ values in the parameter space,
 
 \begin{equation}
 \label{gen-berry1984quantal}
  \displaystyle\int_0^t dt'B_d(t')=\int_{\mathcal C} d{\bf x}.B_d({\bf x}).
 \end{equation}

Equation (7) is a  general expression valid for systems with adiabatic
  driving. If the time dependence is entirely due to the 
   internal dynamics, i.e the system is prepared in the nonequilibrium state and evolves towards the steady state, 
    then one cannot be sure that such an evolution will be adiabatic. In this case,
     separation between dynamic and geometric parts is not possible. So, Eq. (\ref{5}) will not
     be a solution of Eq. (\ref{der-U}).  
     The externally controlled driving allows the evolution to be adiabatic such that Eq. (\ref{5}) is valid
       and hence separation between
       dynamic and geometric parts is possible.

 Assuming 
 the contour  in Eq. (\ref{berryEq}) to be closed (a
 fixed time period) and piecewise smooth, we 
can use Stokes' theorem and 
rewrite the contour integral as a surface integral over the surface $\mathcal S$
 enclosed by the contour. This makes 
the factor
geometric in nature. Note that,  the dynamic part can not be given a geometric interpretation because
 there is no explicit time derivative which can be converted to a parametric integral. 
 Equation (\ref{gen-berry1984quantal}) then 
 becomes,
\begin{equation}
\label{gen-curve}
 \int_{\mathcal C} d{\bf x}.B_d({\bf x})
 =\displaystyle\oint_{\mathcal C} d{\bf x}.B_d({\bf x})=\oiint_{\mathcal S}\big(\nabla\times B_d({\bf x})\big).dS.
\end{equation}
Here, $dS=dxdy$. The surface integrand   
 $\nabla\times B_d({\bf x})$ is equivalent to the geometric curvature where $B_d({\bf x})$
  represents the geometric vector potential. 
  Note that, unlike the case of an isolated quantum dynamics of wavefunction, 
  Eq. (\ref{gen-curve}) cannot be interpreted
   as a phase factor because it affects the probability associated with an observable. 
    Equation (\ref{gen-curve}) represents a geometric contribution to the time evolution of the density matrix.
    In this reduced system dynamics, 
  $\oiint_{\mathcal S}(\nabla\times B_d({\bf x})).dS$ is 
 analogous
  to the Pancharatnam-Berry phase in isolated quantum dynamics and is not a phase as such, but is geometric in nature.
This PB contribution is a direct manifestation of 
   adiabatic and cyclic evolution of (at least) two parameters over a full time period of the drivings. 

  Equation (\ref{gen-berry1984quantal}) is a general expression for the acquired geometric contribution
  due to adiabatic modulation
 of two parameters. If the modulation is periodic, the contour $\mathcal C$ is closed 
in the parameter space
  and the resultant contribution, Eq. (\ref{gen-curve}), is termed as the PB part. 
  For non-periodic driving $\mathcal C$ is an open contour with a geometric interpretation, called
  the ANG part, which we shall discuss in section \ref{NCE}.
  
\section{ Model Calculation}
\label{srl}
\begin{figure}
\centering
\includegraphics[width=7.5cm]{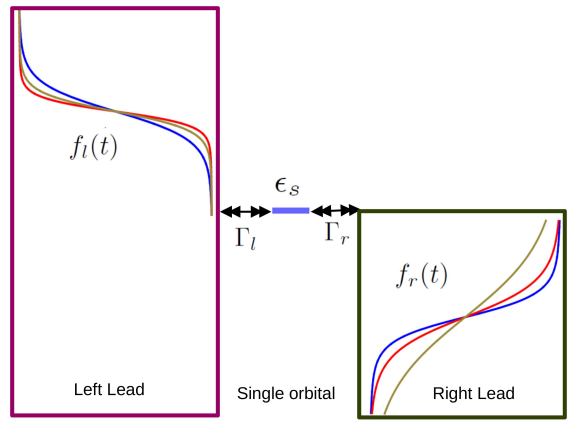}
\caption{Schematics of the driven single resonant level model. The leads are metallic and spinless charge transfer 
 occurs between the single orbital and leads. The Fermi distributions of the metal 
 leads in resonance with the orbital's energy, $\epsilon_s$, are driven 
  periodically in time. $\Gamma_{l(r)}^{}$ is the coupling of the orbital with 
    the left (right) lead.}
\label{scheme}
\end{figure}

We consider a single electronic orbital coupled 
 to two electronic reservoirs kept at different chemical potentials as shown in Fig.(\ref{scheme}).
The Hamiltonian for this system is given by
\begin{align}
\hat{H}&=\hat{H}_s+\hat{H}_l+\hat{H}_r+\hat{H}_{s\nu}\nonumber\\
&=\epsilon_s\hat{c}_s^\dag \hat{c}_s^{}+\displaystyle\sum_l\epsilon_l\hat{c}_l^\dag \hat{c}_l
+\displaystyle\sum_r\epsilon_r\hat{c}^\dag_r
\hat{c}_r\nonumber\\&+\sum_{\nu\in l,r}(T_{s\nu}\hat{c}_s^\dag \hat{c}_\nu+h.c.),
\end{align}
where $s$, $l$, $r$, are the system, the right and the left lead orbitals, respectively. 
$\hat H_{s\nu}$ is the system-lead coupling Hamiltonian such that $T_{s\nu}$ is the
coupling between the system and leads with $\nu=l,r$, and
 $\hat{c}_s^\dag( \hat{c}_s^{})$, $\hat{c}_l^\dag( \hat{c}_l^{})$ and $\hat{c}_r^\dag( \hat{c}_r^{})$
represent the electronic creation(annihilation) operators for the system, left and right leads, respectively.
The leads have no interactions and are always in equilibrium (they act as reservoirs) and exchange spinless charges
 with the system.
  
   We consider the case when the thermodynamic equilibrium of the electronic reservoirs is modulated in time by  
   altering the temperature or the chemical potential.   
    A possible way to experimentally visualize
the driven quantum junction would be to connect the metal contacts (electronic reservoirs)
to a time varying
thermostat. This would allow control on the Fermi-functions of the metal leads
through the time dependent temperatures. Or one may use a time varying gate-voltage
to modulate energy of the orbital or use voltage pulses to control chemical potential of the leads.
All these in turn will change the Fermi distribution of the leads.

 The driven quantum master equation for the reduced system dynamics is given by,
     \begin{eqnarray}
     \label{f-qme}
\dot\rho(t)&=&
 [\Gamma_l\tilde f_l(t)+\Gamma_r\tilde f_r(t)]
 [\hat{c}_s\rho(t)\hat{c}_s^\dag- \hat{c}_s^\dag\hat{c}_s\rho(t)]
 \nonumber\\
&+&[\Gamma_lf_l(t)+\Gamma_rf_r(t)]
[\hat{c}_s^\dag\rho(t)\hat{c}_s-\rho(t)\hat{c}_s\hat{c}_s^\dag].
\end{eqnarray}
 Here, $\Gamma_X=\pi n_X(\epsilon_s)|T_{sX}|^2$ with $n_X$ representing the density of states of 
lead $X$. The adiabatic limit is valid for $t_p(\Gamma_l+\Gamma_r)\gg 1$
and $\tau_s\approx (\Gamma_l+\Gamma_r)^{-1}$. 
The time dependent Fermi-functions for the leads is
 $f_X(t)=(e^{\bar\beta_X(\epsilon_s-\mu_X)}+1)^{-1}$, and $\tilde f_X(t)=1-f_X(t)$. 
 $\mu_X$ and $\bar\beta_X$ is 
 the chemical potential  and inverse temperature of the
 $X\in l,r$-th lead, respectively. The Fermi-functions depend on both temperature and the chemical potential.
Changing either of these
    inherently affects the Fermi functions of the leads. So we choose the driving protocol in terms 
     of the Fermi distributions for simplicity. This implicit dependence
  on time either via the inverse temperatures $\bar\beta_X (t)=(k_BT(t))^{-1}$ or the chemical potential of the 
  $X$th
   lead,
  $\mu_X \rightarrow\mu_X(t)$  is arbitrary as long as the bath correlations die
   fast.
  For the model considered here, the density vector contains only two elements $\rho_{11}$ and $\rho_{00}$
   representing the population of the many-body state with 1 and 0 electrons, respectively. Coherences
    do not couple to populations and  die off exponentially.
     Since we are interested in the steady-state dynamics (adiabatic driving), coherences will be ignored.
 We define
  $|\rho(t)\rangle\rangle
  =\{\rho_{11},\rho_{00}\}$. The time dependent Liouvillian in Eq. (\ref{qLiouville}) is then,
 \begin{eqnarray}
  \label{liouvillian}
\hat{\mathcal{L}}(t)&=&2\begin{pmatrix}
-\alpha(t)&\beta(t)\\
\alpha(t)&-\beta(t)\\
\end{pmatrix},
\end{eqnarray}
 where $\alpha(t)=\Gamma_l\tilde f_l(t)+\Gamma_r\tilde f_r(t)$ and 
 $\beta(t)=\Gamma_lf_l(t)+\Gamma_rf_r(t)$ are the system to leads
 and leads to system electron transfer rates, respectively. 
  
  As we shall discuss below, for the single resonant level case that we consider here, 
  when the Fermi functions are adiabatically modulated (i.e the parameter space, 
   {\bf x}  is composed of
    the time-dependent parameters $f_l(t) $ and $ f_r(t)$),
  the  vector potential, $B_d(f_l,f_r)$ is non-zero.
 However, the overall PB contribution vanishes (Sec.\ref{NCE}).
  This happens because the curvature is zero, $\nabla\times B_d(f_l,f_r)=0$, as a result
   of the periodic driving. So, $B_d(f_l,f_r)$ 
  is a conservative or irrotational field in the parameter space.
 Thus, for
   a two parameter periodic and adiabatic evolution, the PB contribution 
   doesn't affect the dynamics of the reduced density-matrix of the single resonant level, although 
   it does influence the statistics of the electron transfer between the system and leads.

\section{ Electron transfer statistics}
\label{stats}
To quantify
 the effect of PB contribution on the statistics of electron transfer, we
 consider probability distribution function, $P(q,t)$ for the net, $q$, number of electrons
 transferred between system and leads . 
  We define  a generating function $G(\lambda,t)$ corresponding to $P(q,t)$ \cite{uhrmp}
  \begin{equation}
  \label{G-lam}
   G(\lambda,t)=\displaystyle\sum_qP(q,t)e^{\lambda q}\equiv\langle\langle\boldsymbol{1}
   |\rho(\lambda,t)\rangle\rangle.
  \end{equation}
Here $|\boldsymbol{1}\rangle\rangle$ is the identity vector and 
$|\rho(\lambda,t\rangle\rangle$ is the $\lambda$-dependent density vector obeying the equation of motion
(appendix),
\begin{equation}
  \label{M}
   |\dot\rho(\lambda,t)\rangle\rangle=\hat M(\lambda,t)|\rho(\lambda,t)\rangle\rangle.
  \end{equation}
  $\hat{M}(\lambda,t)$ is the $\lambda$-dependent Liouvillian (appendix) given by
  \begin{equation}
 \label{char-L}
\hat M(\lambda,t)=2\begin{pmatrix}
-\alpha(t)&\beta_l(t)e^{\lambda}+\beta_r(t)\\
\alpha_l(t)e^{-\lambda}+\alpha_r(t)&-\beta(t)\\
\end{pmatrix}.\\[2mm]
\end{equation}
Here $\alpha_X=\Gamma_X\tilde f_X(t)$ and $\beta_X=\Gamma_Xf_X(t), X\in l,r$. For $\lambda=0$, Eq. (\ref{M})
reduces to Eq. (\ref{liouvillian}) and $|\rho(\lambda=0,t)\rangle\rangle=|\rho(t)\rangle\rangle$. 

Since we are interested in the steady state fluctuations, we define a scaled cumulant
 generating function\cite{uhrmp,max-uh},
 \begin{equation}
 \label{full S}
  S(\lambda)=\lim_{t \rightarrow\infty}\frac{1}{t}\ln G(\lambda,t).
 \end{equation}

 In the long time limit, it can be shown that 
  scaled cumulant generating function is additively separable into two parts\cite{ren2010berry,sin} (appendix), viz.
  dynamic, $S_d(\lambda)$, and a geometric, $S_g(\lambda)$); 
  i.e $S(\lambda,t)=S_d(\lambda,t)+S_g(\lambda,t)$
  where,
 \begin{eqnarray}
 \label{s-dyn}
S_d(\lambda)&=&\frac{1}{t_p}\displaystyle\int_0^{t_p}dt'\zeta_+(\lambda,t'),\\
\label{s-geo}
S_g(\lambda)&=&\frac{-1}{t_p}\displaystyle\int_0^{t_p}dt'\langle\langle L_+(\lambda,t')
|\frac{\partial}{\partial t'}| R_+(\lambda,t')\rangle\rangle.
\end{eqnarray}
 $|R_+(\lambda,t')\rangle\rangle [\langle\langle L_+(\lambda,t')|]$ denote the instantaneous
 right [left] eigenvectors of $\hat M(\lambda,t')$ corresponding to the instantaneous smaller
 eigenvalue $\zeta_+(\lambda,t')$. The eigenvalues of $\hat{M}(\lambda,t)$ in Eq. (\ref{char-L}) are,
 \begin{small}
 \begin{eqnarray}
 \label{zetaP}
  \zeta_\pm(\lambda,t)&=&-\!\Gamma\!
  \pm\!\sqrt{\Gamma^2\!+\!4(\alpha_r(t)\beta_l(t)(e^\lambda\!-\!1)\!+\!\alpha_l(t)\beta_r(t)(e^{-\lambda}\!-\!1))}
  \nonumber\\
 \end{eqnarray}
\end{small}Here $\Gamma=\Gamma_l+\Gamma_r$. Also, the measurement time,
 $t=\nu t_p$, where $\nu$ is the number of cycles and
  $t_p$ is the time-period of the driving such that $\Gamma t_p\gg 1$.
 We can write $S_g$ as a line integral over a closed contour $\mathcal C$ defined in the parameter space such that,
\begin{eqnarray}
\label{berry1984quantal-t}
 S_g(\lambda)&=&\frac{-1}{t_p}\oint_{\mathcal C} d{\bf x}.\langle\langle L_+({\lambda, \bf x})
 |\partial_{{\bf x}}|R_+({\lambda,
 \bf x})\rangle\rangle.
\end{eqnarray}
Here, vector ${\bf x}$ contains system parameters modulated by the external driving.
 For $\lambda=0$, the integrand in Eq. (\ref{berry1984quantal-t})
is equivalent to $B_d({\bf x})$ in Eq. (\ref{berryEq}).
We can now convert
 the line integral to a surface integral over the contour area, $\mathcal S$, for a closed $\mathcal C$.
  Equation (\ref{berry1984quantal-t}), can be recast as,
\begin{equation}
\label{Berry S}
 S_g(\lambda)=\frac{1}{t_p}\oiint_{\mathcal S}dxdy\mathfrak{B}^\lambda(x,y)
 \end{equation}
 with $\mathfrak{B}^\lambda(x,y)=\mathfrak{B}^\lambda_{xy}-\mathfrak{B}^\lambda_{yx}$ being equivalent to 
 the Pancharatnam-Berry curvature in the parametric  space of $x$ and $y$ (appendix), where
\begin{eqnarray}
\label{BC}
 \mathfrak B_{xy}^\lambda&=&\frac{\langle\langle L(\zeta_-)|\partial_{x}\hat M(\lambda)
 |R(\zeta_+)\rangle\rangle\langle\langle L(\zeta_+)|\partial_{y}\hat M(\lambda) |R(\zeta_-)\rangle\rangle }
 {-(\zeta_+-\zeta_-)^2}.\nonumber\\
\end{eqnarray}
When $\lambda=0$,
   $\mathfrak B^{\lambda=0}(x,y)\equiv \nabla\times B_d(x,y)$.
 Note that, the Pancharatnam-Berry curvature (Eq. (\ref{BC})) is identically zero if only a single parameter 
 is changed
  in periodic manner. This is because $\mathfrak B^\lambda(x,y)=0$. 
  If two mutually dependent parameters are changed, $\mathfrak B_{xy}^\lambda=\mathfrak B_{yx}^\lambda$, resulting
   in vanishing curvature, $\mathfrak B^\lambda(x,y)=0$.

Here we choose to modulate the Fermi functions of the leads, $i.e.$, $x=f_l(t),y=f_r(t)$.
The $\lambda$-dependent PB curvature, $\mathfrak B^\lambda(f_l,f_r)$, simplifies to,
\begin{equation}
\label{b-curve-f}
 \mathfrak B^\lambda(f_l,f_r)=\frac{e^{-\lambda}(e^\lambda-1)^2\Gamma_l\Gamma_r(\Gamma_l-\Gamma_r)}
 {\{\Gamma_l^2+\Gamma_r^2+2\Gamma_l\Gamma_r(\tilde f_l-f_l)(\tilde f_r-f_r)+\mathfrak{Z}_\lambda^{}\}^{\frac{3}{2}}}
 \end{equation}
 with
 \begin{equation}
 \mathfrak{Z}_\lambda^{}= 4\Gamma_l\Gamma_r(f_l\tilde f_r e^\lambda+f_r\tilde f_l e^{-\lambda}).
\end{equation}
When, $\Gamma_l=\Gamma_r$, the PB curvature in Eq. (\ref{b-curve-f}) is zero. This happens because 
$\langle\langle L_+(\lambda,t')|\dot R_+(\lambda,t')\rangle\rangle$ in Eq. (\ref{s-geo}), 
becomes a total time derivative which integrated over the time 
 period $t_p$ becomes zero. So, under a symmetric coupling, $\Gamma_l=\Gamma_r$, the Pancharatnam-Berry
  contribution is zero and the statistics is   
 governed solely by the dynamic part, $S_d(\lambda)$.
 Below we shall always consider the case when
  $\Gamma_l\ne\Gamma_r$, so that we have a finite geometric (PB) contribution.
  From Eq. (\ref{BC}), we also note that when $\lambda=0$, $\mathfrak B^{\lambda=0}(f_l,f_r)=0$. 
  That is the dynamics of the reduced density matrix of the  single resonant level is not affected
  by the geometric (PB) contribution. However the statistics of electron transfer is influenced by
  the geometric part, as we discuss below.

We focus on the $i$th cumulants of the net probability distribution
 function $P(q,t)$ which are obtained from the $i$th $\lambda$-derivatives of the scaled cumulant generating function,
\begin{eqnarray}
 C^{(i)}
 &=&\frac{d^i}{d\lambda^i} S_d(\lambda)\big|_{\lambda=0}+\frac{d^i}{d\lambda^i}S_g(\lambda)\big|_{\lambda=0}.
\end{eqnarray}
Equation (\ref{b-curve-f}) can be substituted in
Eq. (\ref{Berry S}) to compute $S_g(\lambda)$ from which the 
geometric correction to cumulants,
$\frac{d^i}{d\lambda^i} S_g(\lambda)|_{\lambda=0}$,
can be
 evaluated.
We find that the correction to the first cumulant (average flux) is zero as reported earlier \cite{hayakawa}. 
The PB contribution has no effect on the average electronic flux between the system and leads.
\begin{equation}
C^{(1)}_g=\frac{1}{t_p}\displaystyle\oiint_{\mathcal S}\bigg(
\frac{d}{d\lambda}\mathfrak B^\lambda(f_l,f_r)\big|_{\lambda=0}\bigg)df_ldf_r=0.
\end{equation}
    Note that the average flux is $j=Tr\{\hat I\rho(t)\}$, where $\hat I$ is current
     operator. Since $|\rho(t)\rangle\rangle$ doesn't have a PB part, $j$ is also independent
      of it. In fact, the expectation value of all single time observables will be
      unaffected from the  PB contribution. 
  The PB part, however, contributes to higher cumulants through higher order
   correlation functions of time dependent observables. The  contributions to fluctuation (second cumulant) 
and skewness (third cumulant) from the geometric parts are calculated as,
\begin{align}
C^{(2)}_g&=\frac{1}{t_p}\displaystyle\oiint_{\mathcal S} 
\frac{d^2}{d\lambda^2}\mathfrak B^\lambda(f_l,f_r)\big|_{\lambda=0}
 df_l~df_r \\
 \label{fluc}
 &=\frac{1}{t_p}\frac{2C_A\Gamma_l\Gamma_r(\Gamma_l-\Gamma_r)}{(\Gamma_l+\Gamma_r)^3},\\
 C^{(3)}_g&=\frac{1}{t_p}\displaystyle\oiint_{\mathcal S} \bigg(\frac{d^3}{d\lambda^3}
 \mathfrak B^\lambda(f_l,f_r)\big|_{\lambda=0}
 df_l~df_r\bigg)\\
 \label{skew}
 &=\frac{1}{t_p}\frac{36\Gamma_l^2\Gamma_r^2(\Gamma_l-\Gamma_r)}{(\Gamma_l+\Gamma_r)^5}
 \displaystyle\oiint_{\mathcal S}(f_l-f_r)df_ldf_r.
\end{align}
Here $C_A=\oiint_{\mathcal S}df_ldf_r$ is the contour area  in the parameter space
 of $f_l$ and $f_r$, $0\le f_l,f_r\le 1$. We give analytical expressions of $C_A$ for
  a sinusoidal driving in the appendix. 
Therefore, the PB contribution has a quantitative effect on the statistics of electron transfer through
 the  second and higher order cumulants. The PB corrections to the second and third cumulants
  as given in Eqs. (\ref{fluc}) and (\ref{skew}), 
  can be positive or negative depending on the relative values of $\Gamma_l$ and $\Gamma_r$.

 The statistics
 of electron transfer is usually quantified using the Fano-factor ($F$) \cite{Fano01,cox,Fano5}
and is defined as the ratio between the second and first cumulants. 
When $F>1 (F<1)$, the transferred electrons between system and leads are correlated (anti-correlated)
and gives rise
to bunched
\citep{silverman,ab1}(antibunched, \citep{Kondo-bunch,zarchin-bunch}) statistics. In the present case, $F$ is
obtained as,
     \begin{align}
   F&=\frac{C_d^{(2)}+C^{(2)}_g}{C_d^{(1)}}\\
   &=\frac{C_d^{(2)}}{C_d^{(1)}}+\frac{2C_A\Gamma_l\Gamma_r(\Gamma_l-\Gamma_r)}
   {t_p(\Gamma_l+\Gamma_r)^3C_d^{(1)}},
  \end{align}
where the first term is due to the dynamic part with

\begin{figure}
\centering
\includegraphics[width=7.5cm]{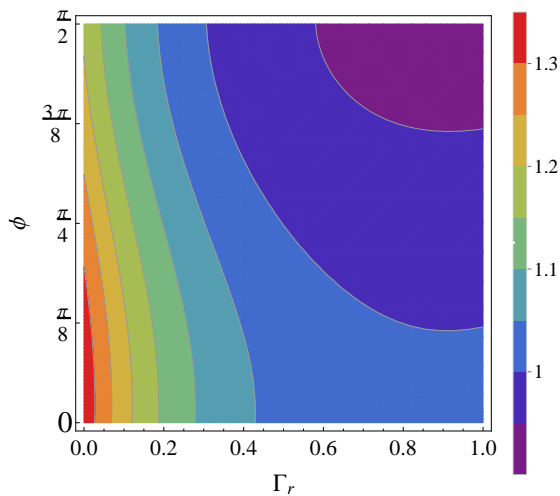}
\caption{The Fano factor ($F$) contour plot for $\Gamma_r$ and $\phi$.
 The PB contribution is introduced by changing the Fermi-functions as
 $f_l(t)=f_l(1-m^2\cos^2(kt)),
 f_r(t)=f_r(1-m^2 \cos^2(kt+\phi))$. $k=\pi/t_p$ is the driving 
  frequency, $t_p$ is the driving time period, $\phi$ is the phase difference between the two drivings and $0<m<1$. 
 Simulation parameters are $ m=0.9, k=1,\nu=10,f_l=0.9,f_r=0.2.$ $\Gamma_l=0.9. $ Second curve from the 
  top-right represent the boundary between $F>1$ and $F<1$.}
\label{Fano}
\end{figure}

\begin{widetext}
 
\begin{eqnarray}
C_d^{(1)}&=&\frac{1}{t_p}\displaystyle\int_0^{t_p}dt
\frac{2[\alpha_r(t) \beta_l(t) - \alpha_l(t) \beta_r(t))]}
{
\sqrt{(\alpha(t)-\beta(t))^2 + 4 (\alpha_r(t) \beta_l(t) + \alpha_l(t) \beta_r(t))}},\\
C_d^{(2)}&=& \frac{2}{t_p}\displaystyle\int_0^{t_p}dt\frac
{(\alpha_r(t) \beta_l(t) ((\alpha(t)-\beta(t))^2 + 2 \alpha_r(t) \beta_l(t)) + \alpha_l(t)\beta_r(t)
[(\alpha(t) - \beta(t))^2 + 6 \alpha_r(t) \beta_l(t)]  + 
 4 \alpha_l^2(t) \beta_r^2(t))}
 {\sqrt{((\alpha(t)-\beta(t))^2 + 4 (\alpha_r(t) \beta_l(t) + \alpha_l(t) \beta_r(t)))}^{3}}.\nonumber\\
 \end{eqnarray}
\end{widetext}
Since $C_d^{(2)}$ is symmetric with respect to interchange of $l$ and $r$,
 while $C^{(2)}_g$ is
 antisymmetric and can be positive,
 negative, or zero, depending on the relative values of $\Gamma_l$ and $\Gamma_r$,
  the statistics of the net electron transfer can be changed by
tuning the fluctuations via PB contribution alone.
Choosing $\Gamma_l>\Gamma_r (\Gamma_l<\Gamma_r)$, 
the PB part enhances (suppresses) the fluctuations.
This effect is shown in Fig.(\ref{Fano}) for sinusoidal drivings:
 $f_l(t)=f_l(1-m^2 \cos^2(kt)), f_r(t)=f_r(1-m^2 \cos^2(kt+\phi))$. Here, $k=\pi/t_p$ is the driving
  frequency, $\phi$ is the phase difference between the two drivings and  $0\le m<1$. 
  Here, 
the quantities $f_l$ and $f_r$ are the Fermi functions
 of the left and the right leads respectively, evaluated 
  at the energy $\epsilon_s$, in the absence of driving ($m=0$).
  As shown in Fig. (\ref{Fano}), the Fano factor can be increased beyond unity 
  by tuning  $\Gamma_r$ or by
   changing the phase difference between the two drivings. For $\phi=0$, 
    the PB contribution is zero (appendix). For nonzero $\phi$, 
    the statistics is bunched for small values of $\Gamma_r$ and tends to become antibunched as $\Gamma_r$ 
    is increased. Over a range of small values of $\phi$, statistics is always bunched.

\begin{figure}
\includegraphics[width=7.5cm]{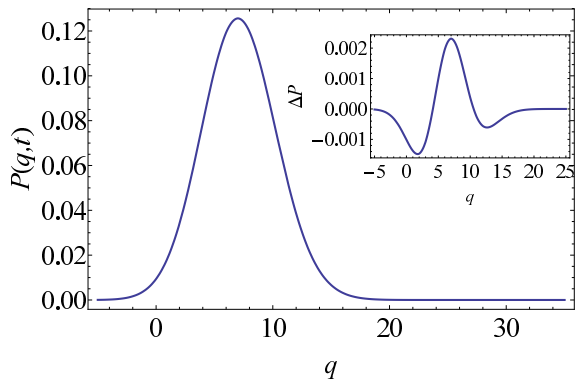}
\caption{The net interpolated PDF in presence of PB contribution ($\phi\ne 0$), 
simulated at $f_l=0.75,f_r=0.25,k=1,\nu=20,k=1,m=0.9,\phi=\frac{\pi}{4}, \Gamma_r=0.25$
and $\Gamma_l=1$.
 The inset shows the difference between the PDFs in presence and absence of PB part.
   }
\label{berry1984quantal-pdf}
\end{figure}

\section{ Steady-state fluctuation theorem}
\label{SSFT}
For a non-driven case, it is known  that
 the generating function satisfies a linear symmetry, $G(\lambda)=G(-\lambda-\mathfrak F)$\cite{uhrmp}, 
 where $\mathfrak F$ is the thermodynamic affinity (nonequilibrium force). 
 This is reflected in the fluctuation symmetry, also referred to as GC symmetry,
 in $P(q,t)$ as\cite{uhrmp},
 \begin{equation}
   \label{ft}
    \lim_{t\rightarrow\infty}\ln \frac{P(q,t)}{P(-q,t)}=q\mathfrak F,
   \end{equation}
where $\mathfrak F=\ln \{f_r(1-f_l)/f_l(1-f_r)\}$ for the resonant level model.

 For the driven case, the full probability distribution function
  is computed by inverting Eq. (\ref{G-lam}) after analytic continuation,
 \begin{equation}
 \label{pdf}
  P(q,t)=\frac{1}{2\pi}\displaystyle\int_0^{2\pi}d\lambda~ G(i\lambda,t)e^{-i\lambda q}.
 \end{equation}
We evaluate Eq. (\ref{pdf}) numerically for a fixed measurement time $t=\nu t_p$. 
In Fig.(\ref{berry1984quantal-pdf}),
  we compare the distributions in presence and in absence of PB contribution. The mean of both
   the distributions is the same but the fluctuations  are different. This is highlighted
   in the inset of Fig.(\ref{berry1984quantal-pdf}), which shows the difference in the values of the PDFs
    in presence and absence of PB part, $\Delta P=P_o(q,t)-P(q,t)$, where $P_o(q,t)$ 
    is the PDF without the PB contribution. 
    Since the average flux is independent of PB, $\int q\Delta P~dq=0$, also due to normalization of 
     $P(q,t)$ and $P_o(q,t)$, $\int\Delta Pdq=0$.

    The distribution, $P(q,t)$, can also be  
   evaluated using the Gartner-Ellis/Varadhan theorem\cite{touchette2009large},  valid at
   large measurement times, where we can write,
   \begin{equation}
\label{ldf-q}
 P(y)\approx N(t) e^{-t\mathfrak{L}(y)},
\end{equation}
where, $y=q/t$ is the rate of electron transfer and $N(t)$ is a time dependent normalization constant. The 
 $"\approx``$ sign indicates that the result is valid only at long times.
$\mathfrak{L}(y)$ is the Legendre-Fenchel transformation of the scaled cumulant generating function
known as the large deviation function\cite{varadhan1966asymptotic,touchette2009large} (LDF)
defined as,
\begin{align}
 \label{legendre}
 \mathfrak{L}(y)&=\ext_\lambda(y\lambda-S(\lambda)),
\end{align}
where $\mbox{ext}$ represents the extremum value(supremum or infimum).
 The full cumulant generating function for the driven case, $S(\lambda)$, is obtained by 
 combining Eqs. (\ref{s-dyn})
and (\ref{Berry S}), where $\zeta_+(\lambda,t)$ and $\mathfrak B^\lambda(f_l,f_r)$ are given by 
Eqs. (\ref{zetaP}) and (\ref{BC}), respectively. 
The large deviation method is based on the saddle point approximation \cite{orszag1978advanced} 
 and is reliable
 only when Eq. (\ref{legendre}) is strictly convex or concave and a well defined 
 extremum value for $\lambda$ 
 exists\cite{touchette2009large}.

 \begin{figure}
\includegraphics[width=7.5cm]{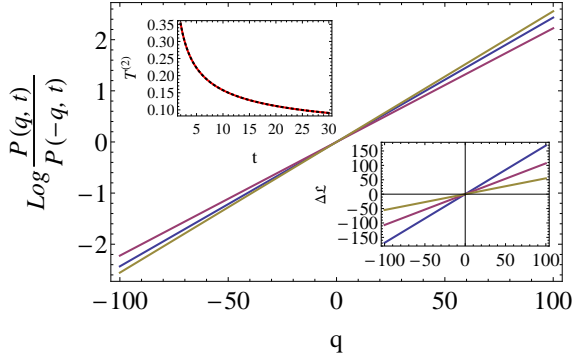}
\caption{The steady state driven fluctuation theorem, Eq.(\ref{driven-ft}).
All curves are simulated at $\Gamma_l=\Gamma_r=0.25,
f_l=0.75,f_r=0.25,k=1,\nu=20,k=1,m=0.9$. The three curves
 with different slopes ($R$) correspond to 
$\phi=0(R=1.38),\frac{\pi}{3}(1.61),\frac{\pi}{4}(1.53)$.
The bottom-right inset shows results for $\Delta\mathfrak{L}$, Eq. (\ref{driven-ldf}),
at $\phi=0,\Gamma_l=0.75,\Gamma_r=0.25$. As $f_l$ increases the slope ($R$) increases, 
$f_l=0.9(R=1.69),0.6(1.07), 0.4 (0.55)$. All other parameters are the same as in the main figure.
The top left inset represents the decay of  Eq. (\ref{T2}), fitted with a 
power law: $t^{-0.49}$ for $q=9$, 
$\phi=\pi/4,m=0.2,\Gamma_l=0.75$ and $\Gamma_r=0.25$ The time axis has been rescaled by a factor 
 of $10^{-2}.$ 
  }
\label{FTplot}
\end{figure}
 Using Eq. (\ref{full S}), 
  in the long time limit, 
     Eq. (\ref{pdf}) can be written as,
      \begin{eqnarray}
      \label{pqt}
       P(q,t)\approx\frac{1}{2\pi}\displaystyle\int_0^{2\pi}d\lambda e^{-i\lambda q+tS(i\lambda)}.
      \end{eqnarray}
We expand the function, $i\lambda q-tS(i\lambda)$ in Eq. (\ref{pqt}) 
around its extremum value $\lambda^*_q$ (the saddle point),
and retain the first two
leading order terms to get,
\begin{eqnarray}
 P(q,t)&\approx&e^{-t\mathfrak{L}(y)}T^{(2)}
\end{eqnarray}
where,
\begin{eqnarray}
\label{T2}
 T^{(2)}&=&\frac{1}{2\pi}\displaystyle\int_0^{2\pi} d\lambda\exp\bigg\{
 \frac{-t}{2t_p}\int_0^{t_p}d\tau\frac{d^2\zeta_+(i\lambda,\tau)}{d(i\lambda)^2}(i\lambda-\lambda_q^*)^2\bigg\}\nonumber\\
\end{eqnarray}
 We evaluate the term  $ T^{(2)}$ numerically and find that it asymptotically 
   goes to zero as a power law $1/\sqrt{t}$, as shown in
 the top left inset of Fig. (\ref{FTplot}). At long times, $(1/t)\log T^{(2)}$
  therefore, decays as an inverse power law, $1/t$. 
  Thus, at large measurement times, the cumulant generating function can be approximated using
 the large deviation result, Eq.(\ref{ldf-q}).
     
    \begin{figure}
\centering
\includegraphics[width=7.5cm]{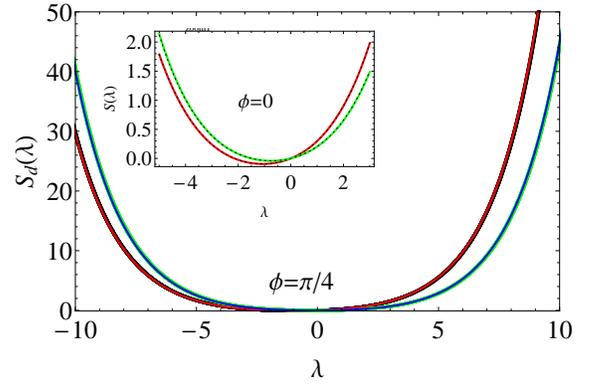}
\caption{The dynamic cumulant generating function, $S_d(\lambda)$ (solid) and linearly shifted,
$S_d(-\lambda-R)$ (dotted). The curves are indistinguishable due to the GC symmetry, 
$S_d(\lambda)=S_d(-\lambda-R)$ where, $R$ is the thermodynamic force given in Eq. (\ref{TF}).
All curves are simulated at $\Gamma_l=0.75, \Gamma_r=0.25,
f_l=0.75,f_r=0.25,k=1,\nu=20,k=1,\phi=\pi/4$. The two curves, red, blue are 
evaluated for
$m=0.2(R=2.153),0.5(1.92)$ respectively.
The inset is evaluated at $\phi=0,\Gamma_l=0.75,\Gamma_r=0.25$. The two curves (red, green) represent
$m=0.2(2.145),0.5 (1.9)$ respectively.
   }
\label{Slambda}
\end{figure}
\begin{figure}
\includegraphics[width=7.5cm]{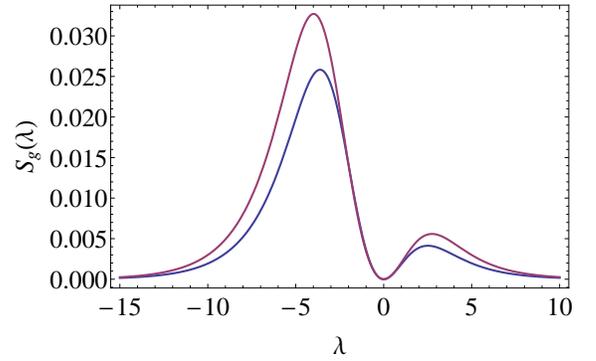}
\caption{The asymmetric geometric cumulant generating function.
$\Gamma_l=1$ (higher peak), 0.75 (lower peak). All other parameters are same as in 
Fig. (\ref{Slambda}). Because of the asymmetric behavior of $S_g(\lambda)$, the linear 
symmetry, $S(\lambda)=S(-\lambda-R)$, does not hold in the presence of the geometric 
contribution, resulting in violation
 of the FT, Eq. (\ref{driven-ft}).
   }
\label{Sgeolambda}
\end{figure}

In the case of heat transfer between two thermal baths, it was reported
by Ren et al\cite{ren2010berry}, that 
the scaled cumulant generating function does not satisfy the usual symmetry, $S(\lambda)=
S(-\lambda-\mathfrak F)$ , even in the absence of the PB contribution. In the present case, however, 
we find that such a symmetry is preserved, as we discuss below.

The LDF  obtained from Eq. (\ref{legendre}) is of the form,
   \begin{eqnarray}
   \label{ldf-lam}
    \mathfrak{L}(q,t)=\frac{q}{t}\lambda^*_q-S(\lambda^*_q),
   \end{eqnarray}
where $\lambda_q^*$ is the value of $\lambda$ at fixed time $t$ which satisfies
 the RHS of Eq.(\ref{legendre}). We numerically evaluate
Eq.(\ref{ldf-lam}) for the case when the PB contribution is zero which 
can be achieved by choosing either $\phi=0$ or $\Gamma_l=\Gamma_r$. 
 We find that the following equality is satisfied,
\begin{eqnarray}
\label{driven-ldf}
 \Delta\mathfrak{L}=\mathfrak{L}(-q,t)-\mathfrak{L}(q,t)=q\frac{R}{t},
\end{eqnarray}
where,
\begin{eqnarray}
\label{TF}
R&=&\displaystyle\ln\frac{\int_0^{t_p}d\tau f_r(\tau)(1-f_l(\tau))}{\int_0^{t_p}d\tau f_l(\tau)(1-f_r(\tau))},
\end{eqnarray}
which reduces to $\mathfrak F$ (Eq. (\ref{ft})) in absence of the drivings. Although
 Eq. (\ref{driven-ldf}) is numerically verified (bottom-right inset of Fig. (\ref{FTplot})), 
 its validity can be justified. When $R=0$, 
   $\Delta\mathfrak L=0 $, giving $P(q,t)=P(-q,t)$ at long time, i.e 
   equilibrium is attained. 
   As the system moves out of equilibrium, $R$ in non-zero and a flux develops in the system. 
   The average flux is given below in Eq. (\ref{avg j}) where $R$ is analytically 
   identified as the thermodynamic force.
 For the non-driven 
  case, Eq. (\ref{driven-ldf}) is always valid and $R\rightarrow \mathfrak F$. $\mathfrak F$ 
  is the thermodynamic
   force for the non driven case. So 
  the basic structure of Eq. (\ref{driven-ldf}) for the driven (without geometric part) and non-driven case
  remain the same. Thus identification of the thermodynamic force by $R$ 
   as given in Eq. (\ref{TF}) is consistent with the definition of equilibrium ($P(q,t)=P(-q,t)$) and the flux,
   both for the driven (without the geometric part) and non-driven system.

Equations (\ref{driven-ldf}) and (\ref{TF}) 
together lead to a driven steady-state fluctuation theorem of the 
type\cite{NESSuh},
\begin{eqnarray}
\label{driven-ft}
 \lim_{t\to\infty}\frac{P(q,t)}{P(-q,t)}=e^{qR},
\end{eqnarray}
preserving the GC type of symmetry in the absence
 of geometric term. As long as $f_l>f_r$,
$R$ is a positive quantity and is the thermodynamic force driving the flux in the system.
In Figs.(\ref{FTplot}) and (\ref{Slambda}), we show the validity
 of the GC symmetry and the steady-state driven fluctuation theorem in absence of the geometric term
  . In Fig. (\ref{Slambda}), we show
   the symmetry, $S(\lambda)=S(-\lambda-R)$, where $R$ is defined
   by Eq. (\ref{TF}).

  For the case when the geometric contribution is nonzero,
 $\mathfrak B^\lambda(f_l,f_r)\ne 0$, Eqs. (\ref{driven-ldf}) and (\ref{driven-ft}) are not valid. This 
  is because $S_g(\lambda)$ does not possess the same linear (translational) symmetry 
  as the dynamic part, $S_d(\lambda)=S_d(-\lambda-R)$.
  Lack of this symmetry in presence of the PB part results in the breaking of the steady state FT.
   We  further observe that the symmetry is broken only near low values of $\lambda$ where the effect
   of PB is most prominent as shown in Fig. (\ref{Sgeolambda}). 
   As $\lambda$ increases $S_g(\lambda)$ goes to zero and the linear symmetry in $S(\lambda)$ is
   recovered for large $\lambda$, implying that, in the limit of large $q$, the FT will be violated, however
   Eqs. (\ref{driven-ldf}) and (\ref{driven-ft}) should be recovered for small $q$ values.

  Note that, adiabatic drivings with $\phi\ne0$ give rise to geometric contributions that results
  in the break-down of FT. However, it is to be emphasized that mere presence of the phase-different 
  drivings doesn't violate 
  the FT, since the geometric contribution may still be zero. For example, in the present case,
  the geometric contribution is zero for all $\phi$ if $\Gamma_l=\Gamma_r$,  and the FT remains valid.

\section{Effect of Noncyclic evolutions}     
\label{NCE}  
   
      Adiabatic non-cyclic geometric (ANG) phases
      \cite{pati,PhysRevLett.60.2339} arise when the adiabatic 
      parametrization takes place  in a noncyclic way, i.e. the curve traced in the parameter space
       is not closed. It 
           has been experimentally observed in the evolution of 
           spatial degrees of freedom of neutrons using interferometry \cite{PhysRevA.72.021602}.
            Equation (\ref{gen-berry1984quantal}) is a general expression for the acquired geometric contribution
             due to adiabatic 
   modulation of two parameters. When the contour $\mathcal C$ is cyclic, we arrived at Eq. (\ref{gen-curve}), 
    the PB contribution. In principle, geometric nature of Eq. (\ref{gen-berry1984quantal}) can be extended
     to cases when $\mathcal C$ is noncyclic and leads to the noncyclic adiabatic 
    contributions which also has a geometric interpretation \cite{mukunda1993quantum,
      sjoqvist2000geometric}.

        Equation (\ref{berryEq}) is equivalent to (appendix):
        \begin{eqnarray}
        \label{rhoR}
        |\rho(t)\rangle\rangle&=&\displaystyle\sum_{m=+,-}a_m(0)|R_m(t)
 \rangle\rangle
 e^{-\int_0^tdt'\chi_m(t')
 -\zeta_m(t')},
        \end{eqnarray}
where $\chi_m(t')=\langle\langle L_m(t')|\dot R_m(t'\rangle\rangle$ is the diagonal 
element ($m=+,-$)
 of the matrix $B_d(t')$ in Eq. (\ref{berryEq}). Equation(\ref{rhoR})
 is simply an expansion of $|\rho(t)\rangle\rangle$ in terms of the right eigenvector of $\hat{\mathcal L}(t)$
 with the initial expansion coefficient $a_m(0)$. Similarly,
  $\langle\langle\rho(t)|$ can be expanded in terms of the left eigenvectors, $\langle\langle L_m(t)$
   such that $\langle\langle\rho(0)|=\sum_{m=\pm}b_m(0)\langle\langle L_m(t)|$. As discussed
    in Sec.(\ref{gc}), we can write,
    \begin{equation}
    \label{new1}
     \displaystyle\int_0^tdt'\chi_m(t)=\int_\mathcal{C}\chi_m({\bf x}).d{\bf x},
    \end{equation}
where $\mathcal C$ is now an open contour in the parameter space, ${\bf x}$.
Note that, unlike for the case of cyclic driving with closed contour $\mathcal C$, 
 for noncyclic evolution, $\int_\mathcal{C}\chi_m({\bf x}).d{\bf x}$ is not 
  gauge invariant as we discuss below. In order to extract a gauge invariant geometric contribution
  we
 project Eq.(\ref{rhoR}) with the initial density vector $\langle\langle\rho(0)|$ to obtain
\begin{eqnarray}
\label{projection}
\langle\langle \rho(0)|\rho(t)\rangle\rangle&=&\sum_{m=\pm}\Xi_m (t)\exp\big\{{\int_0^tdt'\zeta_m(t')}\big\},
\end{eqnarray}
where,
\begin{eqnarray}
\label{ncp}
\Xi_m (t)&=&a_m(0)\langle\langle \rho(0)|R_m(t)\rangle\rangle
 e^{-\int_\mathcal{C}\chi_m({\bf x}).d{\bf x}}.
\end{eqnarray}
 $\Xi_m(t)$ is independent of parametrization of path $\mathcal C$ in the parameter space.
                    $\Xi_m(t)$ is also  invariant under a local gauge transformation,
           $|R_m(t)\rangle\rangle\rightarrow \exp(i\eta(t))|R_m(t)\rangle\rangle$, 
           $\langle\langle L_m(t)|
           \rightarrow \langle\langle L_m(t)|\exp\{-i\eta(t)\}$ where $\eta(t)$ is 
           an arbitrary differentiable function. 
   Both these
          factors together guarantee the  geometric 
          nature\cite{mukunda1993quantum,polavieja1998extending} of $\Xi_m(t)$. 
                  The geometric
           nature of phases during noncyclic evolutions has also been shown by closing 
           the open contour using geodesics
        and
            parallel transport law arguments 
           \cite{PhysRevLett.60.2339}.
  Equation (\ref{ncp}) is the general expression for an adiabatic, noncyclic geometric contribution
  (ANG) which reduces to 
  Eq. (\ref{gen-berry1984quantal}) for a cyclic driving over the closed 
  contour $\mathcal C$. Note that, although, the integral in Eq. (\ref{ncp}) is reparametrization
  independent but it is not gauge invariant and hence not an observable for open 
   contour $\mathcal C$.

For an arbitrary driving, in the long time limit , 
only $m=+$ term dominates. The steady state density matrix is given by $|\rho_s(t)\rangle\rangle=
\{\beta(t)/\Gamma,\alpha(t)/\Gamma\}$ 
and $a_m(0)$ is obtained 
 by solving $|\rho(0)\rangle\rangle=\sum_{m=\pm}a_m(0)|R_m(0)\rangle\rangle$.  So,
  \begin{eqnarray}
  \label{ini}
a_+(0)\langle\langle \rho(0)|R_+(t)\rangle\rangle&=&\frac{\alpha(0)^2+\beta(0)^2}
{\Gamma^2\alpha(0)\alpha(t)}\nonumber\\
&\times&
(\alpha(0)\alpha(t)+\beta(0)\beta(t)^2),\\
  \label{chi}
 \chi_+(t)&=& - \frac{d}{dt}\ln\frac{{\alpha(t)}}{\alpha(0)}, 
\end{eqnarray}

Substituting Eqs.(\ref{ini}) and (\ref{chi}) in (\ref{ncp}), we find
for the generalized ANG contribution
\begin{equation}
\label{+ncp}
 \Xi_+ (t)=\frac{\alpha(0)^2+\beta(0)^2}{\Gamma^2\alpha(t)^2}(\alpha(0)\alpha(t)+\beta(0)\beta(t)),
\end{equation}
 for an arbitrary modulation of the thermodynamic equilibrium of the leads.
For noncyclic driving, $t\ne t_p$, the evolution of the density matrix is influenced by 
the geometric contribution and
 therefore, unlike the cyclic case,
 the flux in the junction is also affected by the geometric (ANG) part. 
 
 Taking a time-derivative in Eq. (\ref{berryEq}), we get,
 \begin{eqnarray}
 \label{splitJ}
   |\dot\rho (t)\rangle\rangle&=&(\hat{\mathcal{L}}_d(t)+\hat{\mathcal{L}}_g(t))|\rho(t)\rangle\rangle
   \end{eqnarray}
   where $\hat{\mathcal{L}}_{d(g)}(t)$ is the dynamic (geometric) Liouvillian given by
   \begin{eqnarray}
   \hat{\mathcal{L}}_d(t)&=&U(t)\Lambda(t)U^{-1}(t)\\
  \hat{\mathcal{L}}_g(t) &=&\dot U(t)U^{-1}(t).
  \end{eqnarray}
 
 The steady state electronic flux between the system and
  left lead is defined as $j(t)=e\langle\langle \hat N|\hat{\mathcal L}^{(l)}(t)|\rho_s(t)\rangle\rangle$
  \cite{uhqme}, where
  $\hat N=\{1,0\}$  and $\hat{\mathcal L}^{(l)}(t)$ is the 
   Liouvillian containing only the terms from the left lead and 
    $|\rho_s(t)\rangle\rangle$ is the steady state density matrix. 
   Following Eq. (\ref{splitJ}), we can split,
   $\hat{\mathcal L}^{(l)}(t)=\hat{\mathcal L}^{(l)}_d(t)+\hat{\mathcal L}^{(l)}_g(t)$, corresponding 
    to the dynamic and the geometric Liouvillians.
   Since there exists a time dependent driving, the steady state is
changing with respect to the external driving. At each instant of driving steady
state is well defined. So the flux, $j(t)$, is time dependent and represents the flux at
    each instant in time during the adiabatic change. The dynamic contribution is given by,
  
  \begin{equation}
  \label{jt srl}
   j_d(t)=I^{SN}_o(f_l(t)-f_r(t)),
  \end{equation}
where $I_{o}^{SN}=2e\Gamma_l\Gamma_r/\Gamma$ is 
 the steady state current in absence of driving in the shot noise limit ($T=0K$). 
The geometric (ANG) contribution is,
 \begin{eqnarray}
 \label{jg}
  j_g(t)&=&e\langle\langle\hat N|\hat{\mathcal L}^{(l)}_g(t)|\rho_s(t)\rangle\rangle\\
  &=&e\displaystyle\sum_{m=\pm}\langle\langle\hat N|\hat{\mathcal L}^{(l)}_g(t)|R_m(0)\rangle\rangle
  \langle\langle L_m(0)|\rho_s(t)\rangle\rangle
  \end{eqnarray}
  Here, in the second line, we have used the resolution of unity in terms of
  $|R_m(t)\rangle\rangle$ and $\langle\langle L_m(t)|$. Since, $\langle\langle L_-(t)|\rho_s(t)\rangle\rangle=0$,
   we can write,
   \begin{eqnarray}
   j_g(t)
  &=&e\langle\langle\hat N|\hat{\mathcal L}^{(l)}_g(t)|R_+(0)\rangle\rangle
  \langle\langle L_+(0)|\rho_s(t)\rangle\rangle.
    \end{eqnarray}
  Using $\langle\langle \rho(0)|=\sum_{m=\pm}b_m(0)\langle\langle L_m(0)|$, we get
  
  \begin{eqnarray}
  \label{jg2}
  j_g(t)&=&\frac{e}{b_+(0)}\langle\langle\hat N|\hat{\mathcal L}^{(l)}_g(t)|R_+(0)
  \rangle\rangle\langle\langle \rho(0)|\rho_s(t)\rangle\rangle,
 \end{eqnarray}
 where, $\langle\langle \rho(0)|\rho_s(t)\rangle\rangle$ is given by Eq. (\ref{projection}) at the steady state 
  and contains information about the ANG contribution as given by Eq. (\ref{+ncp}).
Following Eq. (\ref{jg2}), we can substitute the initial 
and the steady state values and write down  the geometric flux as,
\begin{eqnarray}
 j_g(t)=\frac{e\beta(t)}{\Gamma}\chi_+^{(l)}(t),
\end{eqnarray}
where $\chi_+^{(l)}(t)$ is given by Eq. (\ref{chi}) with 
 only the left lead's contribution. For cyclic driving, $\chi_+(t)=0$, since $t$ is an integral 
  multiple of of $t_p$ giving $j_g(t)=0$.

 We can define average dynamic flux per measurement time, $\mathcal{T}$ as
 \begin{eqnarray}
   \langle j_d\rangle&=&\frac{1}{\mathcal{T}}\displaystyle\int_0^{\mathcal{T}} j_d(t)dt\\
  &=&I_o^{SN}\frac{e^{-R}-1}{\mathcal{T}}\displaystyle\int_0^{\mathcal{T}}f_r(t)(1-f_l(t))dt,
  \label{avg j}
 \end{eqnarray}
where, $R$ is defined by Eq. (\ref{TF}) and acts as thermodynamic force that drives the flux.
For the chosen sinusoidal drivings,
Eq. (\ref{avg j}) becomes,
 \begin{eqnarray}
 \label{qd}
\langle j_d^{}\rangle&=& I_o\big(1-\frac{m^2}{2}\big)+I^{SN}_o\frac{m^2}{n\pi} f(n,\phi),
   \end{eqnarray}
with,
   \begin{eqnarray}
   \label{fracq}
    f(n,\phi)=-\frac{1}{4}[f_l\sin(2n\pi)-f_r\{\sin(2n\pi+2\phi)-\sin(2\phi)\}],\nonumber\\
   \end{eqnarray}
and $I_o=I_o^{SN}(f_l-f_r)$,  is the steady state current in the absence of driving.
Here, we have used, $\mathcal{T}=nt_p$. For the cyclic case, $n$ is an integer which 
represents the number of cycles 
 during the driving and $f(n,\phi)=0$. For the noncyclic case,
  $0<n<1$ and $f(n,\phi)\ne 0$.

\begin{figure}
\centering
\includegraphics[width=7.5cm]{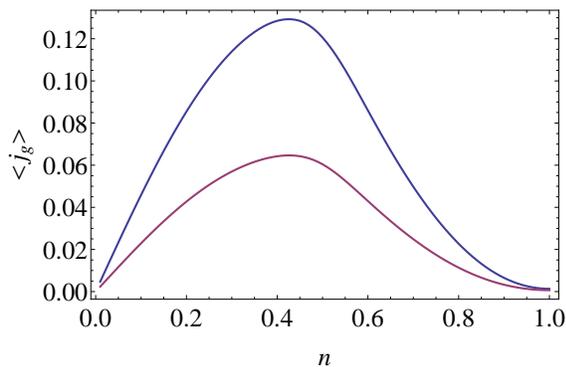}
\caption{The average geometric flux (Eq. (\ref{jgint})) as a function of $n$. The upper (lower) curve
 is simulated for $t_p=10\pi (5\pi)$. It is zero at $n=0,1$. 
  Simulation parameters used are $m=0.9,f_l=0.9,f_r=0.1,\Gamma_l=1.25,\Gamma_r=0.75,\phi=\pi/4$. }
\label{jg plot}
\end{figure}

 The average geometric flux in a measurement window can be written as,
 \begin{eqnarray}
  \langle j_g\rangle&=&\frac{1}{nt_p}\displaystyle\int_0^{nt_p} j_g(t)dt\\
  &=&-\frac{e}{nt_p~\Gamma}\displaystyle\int_0^{nt_p}\frac{\beta(t)\dot\alpha_l(t)}{\alpha_l(t)}dt.
 \end{eqnarray}
For the chosen sinusoidal drivings, we get
 \begin{eqnarray}
 \label{jgint}
  \langle j_g\rangle&=&-\frac{e}{nt_p~\Gamma}\bigg\{\Delta[\beta(nt_p)\log(\tilde f_l(nt_p))]
  +2m^2\Gamma_rf_r\big\{
  \nonumber\\
  &&f_l\cos^2(\phi)\Delta[\tilde f_l(nt_p)\log(f_l(nt_p))]+k \sin(2\phi)A_n\big\}
  \bigg\}\nonumber\\
  \end{eqnarray}
  where,
  \begin{eqnarray}
  A_n&=&\sin(2knt_p)\bigg\{
  4\sqrt{\tilde f_l}\sqrt{\tilde f_l(0)}\arctan\bigg(
  \frac{\sqrt{\tilde f_l\tan(knt_p)}}{\sqrt{\tilde f_l(0)}}
  \bigg)\nonumber\\
  &-&2knt_p(\tilde f_l+\tilde f_l(0))
  -m^2f_l(\log(\tilde f_l(nt_p))-1)
  \bigg\}.
 \end{eqnarray}
Here we have used $\Delta[ X(nt_p)]=X(nt_p)-X(0)$.
 Equation (\ref{jgint}) vanishes for both $n\in \mathbb{Z}$ and $m=0$.
 Note that $\langle j_g\rangle\ne 0$  when $\phi=0$. This tells us that for noncyclic evolutions, 
 even if the system parameters are  
  identically driven, there is always a geometric contribution to the total flux.
 We show the behavior of $\langle j_g\rangle$ in Fig.(\ref{jg plot}) as
 a function of $n$ for two different time-periods. As can be seen, the flux gradually 
  increases from $n=0$ and reaches a maximum and then goes to zero as $n$ approaches 1 (cyclic case).

   \section{ Conclusion} 
   \label{conc}
   We have analyzed the effects of
    cyclic (PB) and noncyclic (ANG) geometric contributions in an adiabatically driven
    current carrying quantum junction. 
     Time evolution of the reduced density matrix for a resonant level is
     shown to be unaffected by the PB contribution arising due to periodic,
     adiabatic driving.  A generating function formalism was used to compute 
     the statistics of the net electron transfered through the junction. 
    The PB contribution to the first cumulant is found to be zero. 
     However the higher order cumulants (fluctuations)
      were affected by the PB contribution. We derived analytic expressions for the 
    geometric contributions to fluctuations. The phase difference between the drivings 
    could be tuned to alter the statistics of the net 
      electron-transfer from antibunched to
       bunched. We also observed that  the  fluctuation theorem 
        (or GC symmetry) is violated
        in presence of a nonzero geometric (PB or ANG) contribution. 
        However, we recover a driven steady-state FT 
        when the geometric contribution vanishes. In this case,
        we identified a thermodynamic force and showed that the GC symmetry is preserved.

          In case of noncyclic evolutions, the geometric contribution (ANG) affects the density matrix 
       of the system which in turn influences the net flux across
       the junction.  Unlike the PB contribution, ANG contribution is non-zero even when the 
       phase difference between the drivings is zero.

     We note that, performing a similar analysis for
        the case when, instead of the Fermi functions, couplings $\Gamma_l$ and $\Gamma_r$
        are modulated periodically in time,
   the PB curvature vanishes altogether in the QME framework, preserving the FT.
   This happens because the
   coupling induced lifetime is neglected in the QME analysis, which is essential to 
    observe geometric effects due to such modulations. Such effects
   are usually incorporated  using  non-equilibrium Greens function technique \cite{hari1}.
     We also observe 
    that in  single electron counting measurements, where one keeps track 
 of only the incoming or outgoing electron transfer processes 
 between the leads and system, the geometric curvature is zero.

    {\em Acknowledgments:} HPG acknowledges the financial support from University Grants Commission, New Delhi
     under the Senior Research Fellowship Scheme. HPG also thanks Ross H. Mckenzie and Hari K. Yadalam
      for interesting discussions.
     BKA thanks the hospitality from Indian Institute of Science, Bangalore.
     UH acknowledges the support from Indian Institute of Science, Bangalore.

\appendix

\section{}
\label{app}
{\bf Derivation of the Pancharatnam-Berry generating function:}
\label{BerryDer}
\label{appsys}
We can re-write Eq. (\ref{f-qme}) as
\begin{eqnarray}
 \frac{\partial\rho(\tau)}{\partial{\tau}}&=&\alpha(\tau)\hat{c}_s\rho(\tau)\hat{c}_s^\dag
-\beta(\tau)\rho(\tau)\hat{c}_s\hat{c}_s^\dag\nonumber\\
&-&\alpha(\tau)\hat{c}_s^\dag\hat{c}_s\rho(\tau)+\beta(\tau)\hat{c}_s^\dag\rho(\tau)\hat{c}_s,
\end{eqnarray}
 where $\alpha(\tau)=\alpha_l(\tau)+\alpha_r(\tau)$ and 
 $\beta(\tau)=\beta_l(\tau)+\beta_r(\tau)$ are the system to leads
 and leads to system electron transfer rates respectively,
\begin{eqnarray}
 \alpha_X(\tau)&=&\Gamma_X(1-f_X(\epsilon_s,\tau)),\\
\beta_X(\tau)&=&\Gamma_X f_X(\epsilon_s,\tau).
\label{rate}
\end{eqnarray}

A quantum
 master equation for such a system is $|\dot\rho(t)\rangle\rangle=\hat{\mathcal L}(t)|\rho(t)\rangle\rangle$, 
$|\rho(t)\rangle\rangle=\{\rho_{11},\rho_{00}\}$ 
is the reduced density vector for the
system containing the population only.
$\hat{\mathcal L}$ is the Liouvillian in the many body space.
We do not include the coherences ($\rho_{01},\rho_{10}$) because 
they exponentially die out and are decoupled from populations.

In this driven case,
 we can obtain the eigenbasis (Eq.(\ref{eigenbasis})) as,
 \begin{eqnarray}
  \Lambda(\tau)&=&\Lambda(0)=\begin{pmatrix}
0&0\\
0&-2(\Gamma_l+\Gamma_r),\\
\end{pmatrix}
 \end{eqnarray}
 and  $U(\tau)$, that diagonalizes $\hat{\mathcal{L}}(t)$, is given by,
 \begin{eqnarray}
  U(\tau)&=&\begin{pmatrix}
\frac{\beta(\tau)}{\alpha(\tau)}&-1\\
1&1\\
\end{pmatrix},
 \end{eqnarray}
The  term, $B_d(\tau)=\text{diag}[U^{-1}(\tau)\dot U(\tau)]$ which when 
 integrated over a time-period, $t=t_p=\pi/k$ is zero.
 Also $\nabla\times B_d(f_l,f_r)=0$. This indicates the field is conservative
 and leaves the dynamics of the density matrix unaffected as it doesn't acquire any geometric contribution 
 during its time evolution.

To quantify the statistics of electron transfer, we start by defining a moment generating
 function, $G(\lambda,t)$ for the PDF corresponding to the net number of particles, $q$ transferred
  between left lead and system.
 The equation of motion for $G(\lambda,t)$ is
\begin{equation}
  \label{M-app}
   \dot G(\lambda,t)=\langle\langle \boldsymbol 1| M(\lambda,t)|\rho(\lambda,t)\rangle\rangle.
  \end{equation}
 $M(\lambda,t)$ is the characteristic counting
  Liouvillian.  
We have  denoted the time and counting-field dependent density vector as
$|\rho(\lambda,t)\rangle\rangle$. We can expand it in the basis
 of the right eigenvector of $M(\lambda,t)$ with time dependent expansion coefficients $a_n(t)$\cite{ren2010berry},
 \begin{equation}
 \label{rev basis}
 |\rho(\lambda,t\rangle\rangle=
 \displaystyle\sum_{n=\pm}a_n(t)e^{\int_0^t\zeta_n(\lambda,t')dt'} |R_n(\lambda,t)\rangle\rangle.                        
                          \end{equation}
                          
The instantaneous eigen values, $\zeta_\pm(\lambda,t)$ of $M(\lambda,t)$ matrix can be written as
\begin{eqnarray}
 \zeta_{\pm}(\lambda,t)&=&-\alpha(t)-\beta(t)\\&\pm&
 \sqrt{\big(\alpha(t)-\beta(t)\big)^2+4(\alpha_\lambda(t)\beta_\lambda(t))}.\nonumber
\end{eqnarray} 

Here $\zeta_+(\zeta_-)$ is the smaller (larger eigenvalue) with 
  $\alpha_\lambda(\tau)=\alpha_l(\tau)e^{-\lambda}+\alpha_r(\tau)$ and $ 
\beta_\lambda(\tau)=\beta_l(\tau)e^{\lambda}+\beta_r(\tau)$. 
Substituting Eq.(\ref{rev basis}) in Eq.(\ref{M}), time evolution of the expansion coefficients
 can be written as,
 we get,
 \begin{align}
  \displaystyle\sum_n\dot a_n(t)&e^{\int_0^t\zeta_n(\lambda,t')dt'}|R_n(\lambda,t)\rangle\rangle\nonumber\\
  &=-\displaystyle\sum_na_n(t)e^{\int_0^t\zeta_n(\lambda,t')dt'}|\dot R_n(\lambda,t)\rangle\rangle.
 \end{align}

Left multiplying by $\langle\langle L_m(\lambda,t|$ and using 
$\langle\langle L_m(\lambda,t|R_n(\lambda,t\rangle\rangle=\delta_{mn}$ gives,
\begin{eqnarray}
  \label{amtf}
 \dot a_m(t)&=&-a_m(t)\langle\langle L_m(\lambda,t)|\dot R_m(\lambda,t)\rangle\rangle\nonumber\\
 &
 -&\displaystyle\sum_{m\ne n}a_n(t)e^{\int_0^t(\zeta_n(\lambda,t')-\zeta_m(\lambda,t'))dt'}\nonumber
 \\
  &\times&
 \langle\langle L_m(\lambda,t)|\dot R_n(\lambda,t\rangle\rangle.
\end{eqnarray}
 Left and right eigenvectors together form an orthonormal set. In the adiabatic limit,
since the eigenstates of the system do not mix, 
the inner product of the time derivative of the right eigen vector and the left eigenvector corresponding 
 to different eigenvalues vanishes, i.e 
 $\langle\langle L_m(\lambda,t)|\dot R_n(\lambda,t)\rangle\rangle=0$.
 The solution of Eq.(\ref{amtf}) can now be written down as,
 \begin{align}
 a_m(t)&=a_m(0)\exp\bigg{\{}-\int_0^tdt'\langle\langle L_m(\lambda,t')|\dot R_m(\lambda,t'\rangle\rangle\bigg{\}}.
 \label{amt}
\end{align}
Substituting Eq. (\ref{amt}) in Eq. (\ref{rev basis}), we get
\begin{align}
 |\rho(\lambda,t\rangle\rangle&=\displaystyle\sum_{m=+,-}a_m(0)|R_m(\lambda,t)
 \rangle\rangle\nonumber\\
 &\times
 \exp\bigg\{-\displaystyle\int_0^tdt'\langle\langle L_m(\lambda,t')|\dot R_m(\lambda,t'\rangle\rangle
 -\zeta_m(\lambda,t')\bigg\}.
\end{align}
The generating function is given by the trace of the counting density-matrix, 
$G(\lambda,t)=\langle\langle {\bf 1}|\rho(\lambda,t)\rangle\rangle$. So,
\begin{align}
G(\lambda,t)&=\displaystyle\sum_{m=+,-}a_m(0)\langle\langle {\bf 1}|R_m(\lambda,t)
 \rangle\rangle\nonumber\\
 &\times
 \exp\bigg\{-\int_0^tdt'\langle\langle L_m(\lambda,t')|\dot R_m(\lambda,t'\rangle\rangle-\zeta_m(\lambda,t)\bigg\},
 \end{align}
 where the left and the right eigenvectors of $M(\lambda)$ are obtained as,
\begin{eqnarray}
|R_{\pm}(\lambda,t)\rangle\rangle&=&\{u_\pm(t),1\}\\
\langle\langle L_{\pm}(\lambda,t)|&=&\frac{1}{u_+(t)-u_-(t)}{}\{\pm 1,\mp u_\mp(t)\},
\label{eigenvectors}
\end{eqnarray}
 with,
 \begin{small}
 \begin{eqnarray}
 \label{Uval}
  u_{\pm}(t)&=&\frac{-(\alpha(t)-\beta(t))\pm\sqrt{(\alpha(t)-\beta(t))^2
  +4\alpha_\lambda(t)\beta_\lambda(t)}}{2\alpha_\lambda(t)}.\nonumber\\
  \end{eqnarray}
\end{small}

 We denote  the time period of evolution by $t_p$ and
 assume that the total measurement time can be expressed as multiple
  of the periodic modulation and write $t=n t_p$, when $n>>1,n\in\Re$. So we get,
 \begin{align}
G(\lambda,t)&=-\displaystyle\sum_{m=+,-}a_m(0)
  \langle\langle {\bf 1}|R_m(\lambda,nt_p)
 \rangle\rangle\nonumber\\
 &\times
 {n\int_0^{t_p}dt'[\langle\langle L_m(\lambda,t')|\dot R_m(\lambda,t'\rangle\rangle-\zeta_m(\lambda,t')]}
 \end{align}
At long times, the contribution from
 the eigenvalue ,$\zeta_-$ is  exponentially suppressed. Hence,
 at large times, 
 \begin{align}
 G(\lambda,t)&\approx a_+(0)\langle\langle \boldsymbol{1}|R_+(\lambda,t)\rangle\rangle\nonumber\\
 &
 e^{\frac{t}{t_p}\int_0^{t_p}(\zeta_+(\lambda,t')
 -\langle\langle L_+(\lambda,t'|\dot R_+(\lambda,t'\rangle\rangle)dt'}.
\end{align}
 
 At the steady state, it is more convenient to work with the scaled cumulant
  generating function defined as 
  \begin{eqnarray}
  \label{cum}
   S(\lambda)&=&\displaystyle\lim_{t \rightarrow\infty}\frac{1}{t}\ln G(\lambda,t)\\
 &=& \lim_{n\rightarrow\infty}\frac{1}{n t_p}
 [\ln a_+(0)\langle\langle \boldsymbol{1}|R_+(\lambda,nt_p)\rangle\rangle]\nonumber\\
 &+&\frac{1}{t_p} \int_0^{t_p}\zeta_+(\lambda,t')dt'\nonumber\\
 &-&\frac{1}{t_p}\int_0^{t_p}\langle\langle L_+(\lambda,t')|\dot R_+(\lambda,t')\rangle\rangle dt'.
  \label{factored}
\end{eqnarray}
The first term in Eq. (\ref{factored}) is constant and goes to zero when $n$ is an integer and
$n \rightarrow\infty$, since $|R_+(\lambda,nt_p)\rangle\rangle=|R_+(\lambda,0)\rangle\rangle$. 
For non integer, $n$, $|R_+(\lambda,nt_p)\ne |R_+(\lambda,0)\rangle\rangle$, the term will survive.

So, for integer values on $n$, the scaled cumulant generating function can be expressed as a sum of
a dynamic ($S_d(\lambda)$)and geometric ($S_g(\lambda)$) scaled cumulant generating functions given by,
\begin{eqnarray}
\label{split}
S(\lambda)&=& S_d(\lambda)+S_g(\lambda),
\end{eqnarray}
and
\begin{eqnarray}
 S_d(\lambda)&=&\frac{1}{t_p} \int_0^{t_p}\zeta_+(\lambda,t')dt',\\
 S_g(\lambda)&=& \frac{-1}{t_p}\int_0^{t_p}\langle\langle L_+(\lambda,t')|\dot R_+(\lambda,t')\rangle\rangle dt'.
\end{eqnarray}
The first and second $\lambda $ derivatives of Eq. (\ref{split}) evaluated at $\lambda=0$ give the flux and
steady state fluctuation.
 Replacing the line integral as a contour integral we recover Eq. (\ref{berry1984quantal-t}).

 We now proceed to derive Eqs. (\ref{Berry S}) and (\ref{BC}), i.e 
 express the Berry potential,
 $\langle\langle L_+({\lambda, \bf x})
 |\partial_{{\bf x}}|R_+({\lambda,
 \bf x})\rangle\rangle$  in terms of the parameter derivatives
  of the counting Liouvillian $M(\lambda)$.  Here, $\bf x$ is a vector
 with two parameters $x$ and $y$.
 For the right eigenvector of $\hat M(\lambda)$ corresponding to the $i$-th eigenvalue ($i=+,-$) we have:
\begin{align}
 \hat M(\lambda) |R_i(\lambda,t\rangle\rangle&=\zeta_i|R_i(\lambda,t)\rangle\rangle\\
 \implies \partial_{y}\hat M(\lambda)|R_i(\lambda,t)\rangle\rangle
 &=\partial_{y}\zeta_i |R_i(\lambda,t)\rangle\rangle\\
 \implies \partial_{y}(\hat M(\lambda)-\zeta_i)|R_i(\lambda,t\rangle\rangle &=(\zeta_i-\hat M(\lambda))|\partial_{y}
 R_i(\lambda,t)\rangle\rangle.
\end{align}

Taking the projection with the left eigenvector $\langle\langle L_j(\lambda,t)|$, we get
\begin{align}
\label{left}
 \langle\langle L_j(\lambda,t)| \partial_{y}R_i(\lambda,t)\rangle\rangle
 =\frac{\langle\langle L_j(\lambda,t)|\partial_{y}\hat M(\lambda)|R_i(\lambda,t)\rangle\rangle}{(\zeta_i-\zeta_j)^2}.
\end{align}
Similarly steps can be done with the left eigenvector and taking projection with $|R_j(\lambda,t)\rangle\rangle$,
we get
\begin{align}
\label{right}
 \langle\langle \partial_{x}L_i(\lambda,t)|R_j(\lambda,t)\rangle\rangle
 =\frac{\langle\langle L_i(\lambda,t)|\partial_{x}\hat M(\lambda)|R_j(\lambda,t)\rangle\rangle}{(\zeta_i-\zeta_j)^2}.
\end{align}
The Pancharatnam-Berry potential in the two parameter vector space ${\bf x}$ is given by
$\langle\langle L_+({\lambda, \bf x})|\partial_{{\bf x}}|R_+({\lambda,
 \bf x})\rangle\rangle$, whose curl gives the curvature, $\mathfrak{B}^\lambda(x,y)$,
\begin{align}
\label{par}
 \mathfrak B^\lambda(x,y)&=\langle\langle \partial_{x}L_+(\lambda)|\partial_{y}R_+(\lambda)\rangle\rangle
 -\langle\langle \partial_{y}L_+(\lambda)|\partial_{x}R_+(\lambda)\rangle\rangle.
\end{align}

Substituting Eq. (\ref{left}) and Eq.(\ref{right}) in Eq. (\ref{par}) and using
$\sum_{i=\pm}|R_i(\lambda,t)\rangle\rangle\langle\langle L_i(\lambda,t)|={\bf 1}$, we can recover Eq. (\ref{BC}).

{\bf Evaluation of Contour area, $C_A$}:

\label{appC}
The parametric dependence on time $\tau$ for the Fermi functions can be recast as an equation of ellipse. Let 
$f_l(\tau)$ and $f_r(\tau)$ represent the time dependent Fermi-functions such that
\begin{eqnarray}
 f_l(\tau)&=&f_l(1-m^2\cos^2(k\tau))\\
 f_r(\tau)&=&f_r(1-m^2\cos^2(k\tau+\phi)).
\end{eqnarray}
Here, $0<m<1$, and $\phi$ is the phase difference. These two equations can be recast as a single ellipse
equation of the form, $A f_l(\tau)^2+Bf_r^2(\tau)+Cf_l(\tau)f_r(\tau)+D f_l(\tau)+E f_r(\tau)+F=0$, where,
\begin{eqnarray}
 A&=&\frac{1}{f_l^2},\\
 B&=&\frac{1}{f_r^2},\\
 C&=&\frac{-2\cos(2\phi)}{f_lf_r},\\
 D&=&\frac{(m^2-2)}{f_l}(\sin^2(2\phi)-2\sin^2(\phi)\cos(2\phi)),\\
 E&=&\frac{m^2-2}{f_r}\sin^2(\phi),\\
 F&=&(1-m^2)\sin^2(2\phi)+(2-m^2)^2\sin^4(\phi).
\end{eqnarray}
Here, $C^2<4AB$ preserving the ellipse at all times. The centers of the ellipse is at  $\{x_o,y_o\}$ given by
\begin{eqnarray}
 x_o&=&\frac{EC-2BD}{4AB-C^2},\\
 y_o&=&\frac{DC-2AE}{4AB-C^2}.
\end{eqnarray}
The major ($a_M$) and minor ($a_m$) 
axes are given by
\begin{eqnarray}
 a_M&=&\frac{-F_o}{A\cos^2(\theta_R)+B\sin^2(\theta_R)-0.5\sin^2(\theta_R)},\\
 a_m&=&\frac{-F_o}{A\sin^2(\theta_R)+B\cos^2(\theta_R)-0.5\sin^2(\theta_R)},
\end{eqnarray}
where
\begin{eqnarray}
 F_o&=&Cx_oy_o+Dx_o+Ey_o+Ax_o^2+By_o^2+F,
\end{eqnarray}
and $\theta_R$ is the angle of rotation of the ellipse given by
\begin{equation}
 \theta_R=\frac{1}{2}\arctan\big(\frac{C}{A-B}\big).
\end{equation}

The contour area is given by $C_A=\oiint f_lf_r=\pi a_M a_m.$ For $\phi=0$, $D=E=F=0$  and as a consequence, 
$C_A=0$.

\bibliography{BerrySRL.bib}

\end{document}